\documentclass[twocolumn,prb]{revtex4-1}
\usepackage{amsfonts}
\usepackage{amsmath}
\usepackage{amssymb}
\usepackage[pdftex]{graphicx}
\begin{document}

\title{Proximity fingerprint of $s_{\pm}$ superconductivity}
\author{A. E. Koshelev  and V. Stanev}
\affiliation{
 Materials Science Division, Argonne National
Laboratory, Argonne, Illinois 60439 }

\begin{abstract}
We suggest a straightforward and unambiguous test to
identify possible opposite signs of superconducting order parameter
in different bands proposed for iron-based superconductors
($s_{\pm}$-state). We consider proximity effect in a weakly coupled
sandwich composed of a $s_{\pm} $-superconductor and thin layer of
$s$-wave superconductor. In such system the $s$-wave order parameter
is coupled differently with different $s_{\pm} $-gaps and it
typically aligns with one of these gaps. This forces the other
$s_{\pm}$-gap to be anti-aligned with the $s$-wave gap. In such
situation the aligned band induces a peak in the $s$-wave density of
states (DoS), while the anti-aligned band induces a dip. Observation
of such contact-induced negative feature in the $s$-wave DoS would
provide a definite proof for $s_{\pm}$-superconductivity.
\end{abstract}
\maketitle

Report of superconductivity at T$_c = 26$ K in fluorine-doped
LaFeAsO \cite{LaOFeAs} followed by discovery of several new classes
of iron-based superconductors \cite{RenSmCPL08,122,111,11} with
transition temperatures up to 56 K generated enormous interest in
the condensed-matter community, see reviews
\cite{physicaC,NewJPhys,PaglioneGreene10}.
In spite of extensive research, the symmetry of the order parameter
in these materials remains prominent unresolved issue. The Fermi
surface of the iron-based superconductors is composed of several
electron and hole sheets. Theory strongly suggests that Cooper
pairing in these materials has electronic origin and the
superconducting order parameter has opposite signs in the electron
and hole bands ($s_{\pm}$-state)\cite{Mazin,Kuroki,Graser,Vlad}.
Experimental evidence for this state, however, remains rather
limited.

The ARPES measurements \cite{Wray,Ding,Zhang} indicate full gaps on
both hole and electron bands, at least in some compounds, but they
can not probe the relative signs of the gaps. At present, the
strongest support in favor of the sign-changing order parameters is
coming from the observation of a resonance in the spin excitation
spectrum developing below the superconducting transition
temperature. This resonance has been observed by the inelastic
neutron scattering in most iron-based superconducting
compounds.\cite{MagMode,ChiPRL09,InosovNatPhys09,QiuPRL09,MookPRL10,ShamotoPRB10}
However, the straightforward interpretation of these experiments was
questioned in Ref.\ \cite{OnariPRB10}.

Several recent experiments provide substantial indirect support in
favor of $s_{\pm}$-state. In particular, behavior of quasiparticle
interference with magnetic field in Fe(Se,Te) compound probed by STM
is consistent with this state \cite{Hanaguri10}. %
Also, observation of the microscopic coexistence of
superconductivity and spin density wave in some iron pnictides is
most naturally explained assuming opposite signs of the order
parameters in the electron and hole bands \cite{Fernandes10}.

The most convincing demonstration of $s_{\pm}$-state could come from
phase-sensitive experiments. These kinds of experiments
\cite{Tsuei,VanHarlingen} played a decisive role in convincing
superconductivity community that the order parameter in the cuprate
superconductors has d-wave symmetry.  Even though the theoretical proposals for
similar experiments have been made for iron-based superconductors
\cite{WuPhillips,PhaseSensFe}, they has not been realized yet. The only
phase-sensitive experiment reported so far is the observation of half-integer
flux-quantum jumps of magnetic flux through the loop formed by niobium and
polycrystalline iron-pnictide sample \cite{FluxJumpsNaturePhys10}.  It is
desirable, however, to design an experiment with better control and more
predictable outcome.

\begin{figure}[ptb]
\begin{center}
\centerline{\includegraphics[width=3.3in]{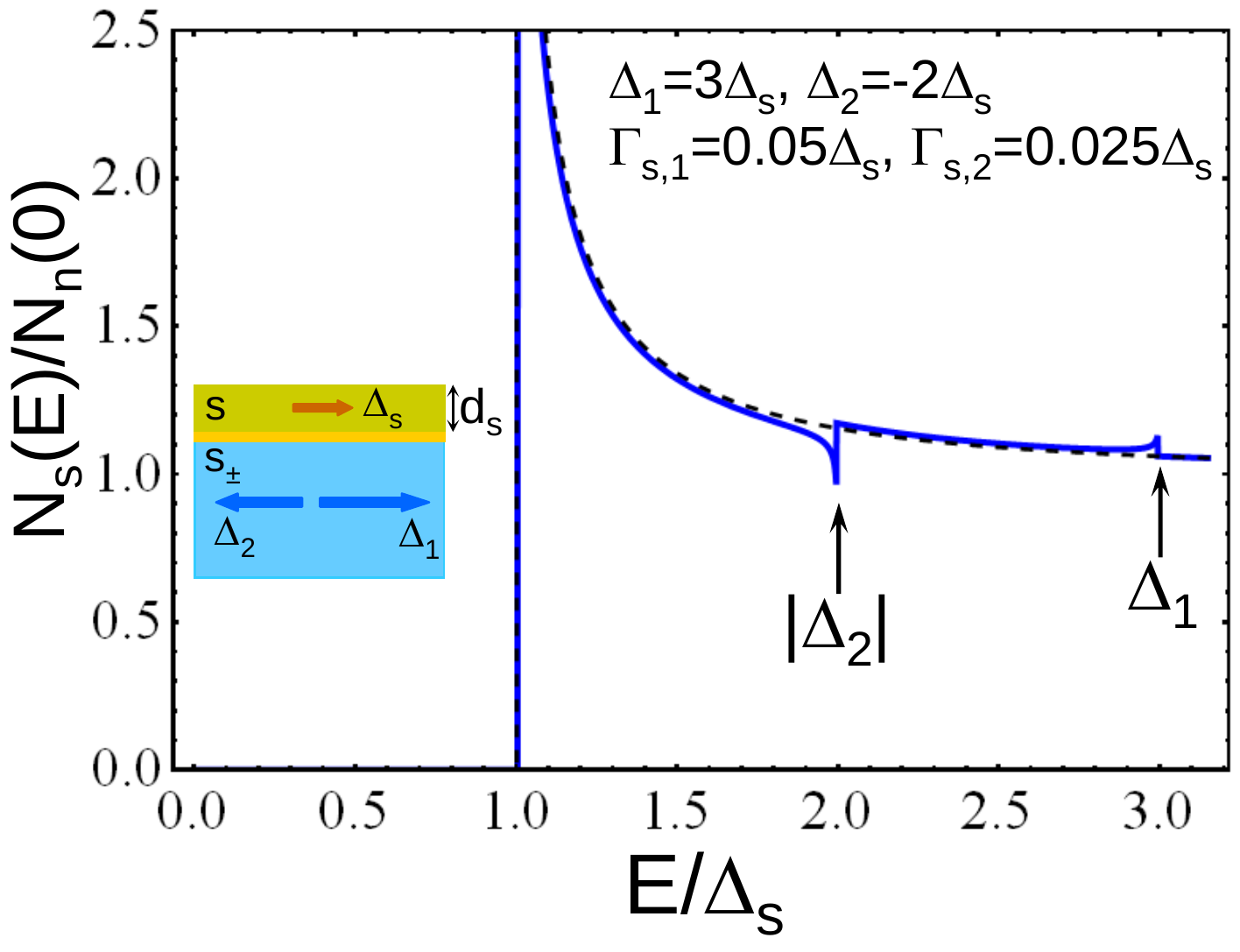}}
\end{center}
\caption{Representative density of states for a thin layer of
$s$-wave superconductor on the top of $s_{\pm}$-superconductor as
illustrated in the inset. The dashed line shows the bulk $s$-wave
DoS. The dip caused by the anti-aligned band provides a definite
fingerprint of $s_{\pm}$-superconductivity.}
\label{Fig-ProxFingerprint}
\end{figure}

In this letter we propose an alternative straightforward test for
relative sign of the order parameter in the electron and hole bands
in the case when the absolute values of the gaps are different. We
consider sandwich composed of $s_{\pm}$ and $s$-wave
superconductors, see inset in Fig. \ref{Fig-ProxFingerprint}.
Peculiar properties of $s$-$s_{\pm}$ Josephson junctions and point
contacts were recognized and studied in several theoretical papers
\cite{Ng,Linder,Chen,Tsai,VS,Ota,Berg}. Nevertheless, to our
knowledge, the proximity effect we describe here was never
mentioned.

For illustration of the effect we consider the simplest situation when
thickness of $s$-wave superconductor is small compared with the coherence
lengths, and coupling between the superconductors is weak. We also assume a
simple two-band model for the $s_{\pm}$ superconductor and a dirty limit for
both materials. While these assumptions allow for a simple analytical treatment
of the problem, none of them is actually essential for the proposed effect.

The first microscopic description of proximity sandwich composed of
two thin superconductors was elaborated by McMillan.\cite{McMillan}
The modern treatment of this problem in dirty limit is based on
Usadel equations \cite{Usadel,Belzig99} which were generalized for
multiband case in Ref.\ \cite{Golubov2}.
Advantage of this approach with respect to more microscopic models
is that it describes properties of junctions via minimum number of
most relevant and physically transparent parameters.

For $s$-superconductor located within $0<x<d_{s}$ the Green's
function $\Phi_{s}$ obeys
\begin{equation}
\frac{D_{s}}{2\omega G_{s}}\left[  G_{s}^{2}\Phi_{s}^{\prime}\right]
^{\prime}-\Phi_{s}=-\Delta_{s},\
G_{s}=\frac{\omega}{\sqrt{\omega^{2}+\Phi _{s}^{2}}},
\label{Green-s}
\end{equation}
where $D_{s}$ is the diffusivity and $\omega=2\pi(n+1/2)T$ are the
Matsubara frequencies. We will be interested only in density of
states of the $s$-wave material and will not need equations for the
$s_{\pm}$ Green's functions.

The boundary condition for the top boundary is simple, $\Phi
_{s}^{\prime}=0$ at $x=d_{s}$. The boundary conditions for the
contact of two dirty superconductors have been derived in Ref.
\cite{KL} and generalized to multiband case in Ref.\
\cite{Golubov2},
\begin{eqnarray}
\xi_{s}G_{s}^{2}\Phi_{s}^{\prime}  &
=&\sum_{\alpha}\frac{\xi_{\alpha}}
{\gamma_{\alpha}}G_{\alpha}^{2}\Phi_{\alpha}^{\prime}\hbox{,\ with }
\gamma_{\alpha}=\frac{\rho_{\alpha}\xi_{\alpha}}{\rho_{s}\xi_{s}
},\label{Bound-curr}\\
\gamma_{B\alpha}\xi_{\alpha}G_{\alpha}\Phi_{\alpha}^{\prime}  &
=&G_{s} (\Phi_{s}-\Phi_{\alpha})\hbox{, with
}\gamma_{B\alpha}=\frac{R_{B}^{\alpha} }{\rho_{\alpha}\xi_{\alpha}}.
\label{BoundResist}
\end{eqnarray}
Here $\rho_{s}$ is resistivity of the $s$-wave superconductor, $\alpha=1,2$ is
the band index, $\Phi_{\alpha}$ and $G_{\alpha}=\omega/\sqrt{\omega^{2}
+\Phi_{\alpha}^{2}}$ are the $s_{\pm}$ Green's functions, $\rho_{\alpha}$ are
partial resistivities for the bands of the $s_{\pm}$ superconductor
\cite{NoteResistDiff}, $R_{B}^{\alpha}$\ are the partial resistances of the
boundary which determine electrical coupling between $s$-wave superconductor
and $s_{\pm}$ bands. In the case of weak coupling, $\gamma_{B\alpha}\gg1$, we
can use approximations $\Phi _{s}\approx\Delta_{s}$ and
$\Phi_{\alpha}\approx\Delta_{\alpha}$ and obtain the approximate boundary
conditions for the $s$-wave Green's function
\begin{equation}
\xi_{s}G_{s}\Phi_{s}^{\prime}=\sum_{\alpha}\frac{G_{\alpha}}{\tilde{\gamma
}_{B\alpha}}(\Delta_{s}-\Delta_{\alpha}) \label{BoundCodAppr2}
\end{equation}
with
$\tilde{\gamma}_{B\alpha}=\gamma_{\alpha}\gamma_{B\alpha}=R_{B}^{\alpha
}/\rho_{s}\xi_{s}$. This condition together with Eq.\
(\ref{Green-s}) allows us to obtain correction to the $s$-wave
Green's function imposed by the contact with
$s_{\pm}$-superconductor. In general, the gap values $\Delta_s$ and
$\Delta_{\alpha}$ have to be found self-consistently but in the case
of weak coupling they can be well approximated by the bulk gaps,
which we assume to be known.

In the case of thin layer, $d_{s}\ll\xi_{s}$, we can expand the
Green's functions,
$\Phi_{s}(x)\approx\bar{\Phi}_{s}+(a_{s}/2)(x-d_{s})^{2}$, where the
parameters $a_{s}$ and $\Phi_{s}$ can be related by Eq.\
(\ref{Green-s})
\begin{equation}
\frac{D_{s}}{2\omega}
G_{s}a_{s}\approx\bar{\Phi}_{s}-\bar{\Delta}_{s}.
\end{equation}
Matching at $x=0$ using Eq.\ (\ref{BoundCodAppr2}) gives
\begin{equation}
\xi_{s}G_{s}a_{s}d_{s}=-\sum_{\alpha}\frac{G_{\alpha}}{\tilde{\gamma}
_{B\alpha}}(\Delta_{s}-\Delta_{\alpha}).
\end{equation}
Solving the last two equations, we obtain
\begin{equation}
\bar{\Phi}_{s}-\bar{\Delta}_{s}\approx-\sum_{\alpha}\frac{\Gamma
_{s,\alpha}(\Delta_{s}-\Delta_{\alpha})}{\sqrt{\omega^{2}+\Delta_{\alpha}
^{2}}} \label{Green-sResult}
\end{equation}
where, following Ref.\ \cite{McMillan}, we introduced the coupling
parameters
\begin{equation}
\Gamma_{s,\alpha}\equiv\frac{D_{s}}{2\xi_{s}d_{s}\tilde{\gamma}_{B\alpha}
}=\frac{\rho_{s}D_{s}}{2d_{s}R_{B}^{\alpha}}=\frac{1}{2e^{2}\nu
R_{B}^{\alpha }d_{s}}
\end{equation}
which have dimensionality of energy. This correction is similar to
the McMillan result \cite{McMillan} for a single-band $s$-wave
superconductors in the linear with respect to $\Gamma_{s,\alpha}$
order.

The density of states of the $s$-wave superconductor is given by
\begin{equation}
N_{s}(E)=\mathrm{Re}\left[
\frac{E}{\sqrt{E^{2}-\Phi_{s}^{2}}}\right].
\end{equation}
Performing analytic continuation of Eq.\ (\ref{Green-sResult}),
$i\omega \rightarrow E-i\delta$,
\begin{equation}
\bar{\Phi}_{s}\approx\bar{\Delta}_{s}+\sum_{\alpha}\frac{\Gamma
_{s,\alpha}\left(  \Delta_{\alpha}-\Delta_{s}\right)
}{\sqrt{\Delta_{\alpha }^{2}-E^{2}}},
\end{equation}
and expanding, we finally obtain
\begin{eqnarray}
N_{s}(E)\!&=&\mathrm{Re}\left[ \frac{E}{\sqrt{E^{2}-\Delta_{s}^{2}}
}\right.  \nonumber\\
& +&\left.\frac{E\Delta_{s}}{\left(  E^{2}-\Delta_{s}^{2}\right)
^{3/2} }\sum_{\alpha}\frac{\Gamma_{s,\alpha}\left(
\Delta_{\alpha}-\Delta _{s}\right)
}{\sqrt{\Delta_{\alpha}^{2}-E^{2}}}\right] .\label{DoSResult}
\end{eqnarray}
This result is valid for any gap parameters.

To proceed further we have to make assumptions about the gap
magnitudes and their signs. We assume that
$|\Delta_{1}|>|\Delta_{2}|>|\Delta_{s}|$, $\Delta_{1}>0$,
$\Delta_{2}=-|\Delta_{2}|<0$. The sign of $\Delta_{s}$ marks
alignment with one of the $s_{\pm}$ bands and it is determined by
the relation between the corresponding Josephson couplings. The
partial Josephson coupling energy between the $s$-wave
superconductor and $\alpha$ band, $E_{J,\alpha}$, is proportional to
the corresponding Josephson current density, $j_{J,\alpha}$, which
is given by the same formula as for a tunneling contact between two
different superconductors, see, e.g., Ref.\ \cite{BaroneBook},
\begin{equation}
E_{J,\alpha}\propto j_{J,\alpha} =\frac{2\hbar}{e
R_B^\alpha}\frac{|\Delta_\alpha||\Delta_s|}{|\Delta_\alpha|+|\Delta_s|}K\left(\frac
{|\Delta_\alpha|-|\Delta_s|}{|\Delta_\alpha|+|\Delta_s|}\right)
\end{equation}
for $T\ll T_c^s$. Here $K(k)=\int_0^{\pi/2}(1-k^2\sin^2x)^{-1/2}dx$
is the complete elliptic integral of the first kind. In the case of
strong inequalities $|\Delta_{s}|\ll |\Delta_{\alpha}|$ we have
$E_{J,\alpha}\propto
(\Delta_s/R_B^{\alpha})\ln(4|\Delta_{\alpha}|/|\Delta_{s}|)$ meaning
that the ratio of the coupling energies is mostly determined by the
ratio of the partial resistivities and only weakly depends on the
gap magnitudes $|\Delta_{\alpha}|$.

For definiteness, we assume that $E_{J,1}>E_{J,2}$ and
$\Delta_{s}>0$ is aligned with the larger gap $\Delta_{1}$, as it is
illustrated in the inset of Fig. \ref{Fig-ProxFingerprint}. In this
case we can rewrite the proximity correction to the DoS as
\begin{eqnarray}
\delta N_{s}(E)&=&\frac{E\Delta_{s}} {\left(
E^{2}-\Delta_{s}^{2}\right) ^{3/2}}  \left[ \frac{\Gamma_{s,1}\left(
\Delta_{1}\!-\!\Delta_{s}\right)
}{\sqrt{\Delta_{1}^{2}-E^{2}}}\Theta(\Delta_{1}\!-\!E) \right.\nonumber \\
&- &\left. \!\frac{\Gamma_{s,2}\left( |\Delta_{2}|\! +\!\Delta_{s}\right)
}{\sqrt{\Delta_{2}^{2}-E^{2}} }\Theta(|\Delta_{2}|\!-\!E)\right],
\label{deltaDoS}
\end{eqnarray}
where $\Theta(x)$ is the step function.  We immediately see that the
\emph{aligned} band induces \emph{positive} correction and the
\emph{anti-aligned} band induces \emph{negative} correction. While
the positive correction is a standard feature of proximity between
two superconductors \cite{McMillan}, the negative anomaly is unique
to $s/s_{\pm}$ proximity. The amplitude of the peak is proportional
to the gap difference $\Delta_1-\Delta_s$, while the amplitude of
the dip is proportional to the gap sum $|\Delta_2|+\Delta_s$. An
example of the $s$-wave DoS for representative parameters is shown
in Fig.\ \ref{Fig-ProxFingerprint}. We also can see that the
$s_{\pm}$ superconductor can both enhance and suppress the $s$-wave
DoS  at energies $E\sim \Delta_s$; the sign of the total correction
in this energy range is determined by the sign of the combination
$\Gamma_{s,1}\sqrt{\frac{\Delta_{1}-\Delta_{s}}{\Delta_{1}
+\Delta_{s}}}-\Gamma_{s,2}\sqrt{\frac{|\Delta_{2}|+\Delta_{s}}{|\Delta
_{2}|-\Delta_{s}}}$. %
The simple analytical result (\ref{deltaDoS}) is obtained in the
linear order with respect to the coupling parameters
$\Gamma_{s,\alpha}$ and does not describe energy region close to the
gap values $|E-\Delta_\alpha|\sim
\Gamma_{s,\alpha}^2/\Delta_\alpha$. In particular, vanishing of the
correction at energies larger than the corresponding gap  leading to
very asymmetric shape of the correction is not an exact result but
just a consequence of this linear approximation.

The $s$-wave DoS can be experimentally accessed in a standard way by
measuring tunneling conductance from the top surface of the sandwich
using scanning tunneling microscopy, point contacts, or making a
planar tunnel junction.
The proposed test only works if there is a noticeable difference
between the absolute values of $s_{\pm}$ gaps so that their features
are sufficiently separated in energy (voltage). %
A good possible choice for the $s$-wave material may be amorphous thin films,
such as Mo$_{x}$Ge$_{1-x}$, because, due to completely incoherent tunneling,
these materials are expected to have comparable coupling with all bands of the
$s_{\pm}$ superconductor. The film thickness has to be smaller or at least
comparable with the coherence length. In a real experiment the anomalies
induced by the $s_{\pm}$ gaps are expected to be less sharp than in the
illustrated ideal situation. They will be smeared, e.g., by temperature, finite
transparency of the interface, and pair-breaking scattering. The optimum
coupling strength between the superconductors for the observation of the effect
has to be in the intermediate range: it should not be too weak so that the DoS
corrections do not vanish in the noise but it should be also not too strong so
that the gaps at the surface are close to the bulk values. This corresponds to
the coupling parameters in the range $\Gamma_{s,\alpha}/\Delta_s=0.01-0.1$.

In summary, we proposed a straightforward test for the $s_{\pm}$
superconducting state using the proximity-induced correction to the
density of state of a conventional superconductor. Coupling between
$s$- and $s_{\pm}$- superconductors typically aligns $s$-wave gap
with one of the $s_{\pm}$-gaps. In this case the anti-aligned gap
induces a \emph{negative} correction to the $s$-wave DoS which which
can serve as a definite fingerprint for the $s_{\pm}$ state. We
analytically evaluated the DoS corrections in the linear order with
respect to the coupling strength.

The authors would like to thank Th. Proslier for useful discussions.
This work was supported by UChicago Argonne, LLC, operator of
Argonne National Laboratory, a U.S. Department of Energy Office of
Science laboratory, operated under contract No. DE-AC02-06CH11357,
and by the \textquotedblleft Center for Emergent
Superconductivity\textquotedblright, an Energy Frontier Research
Center funded by the U.S. Department of Energy, Office of Science,
Office of Basic Energy Sciences under Award Number DE-AC0298CH1088.


\begin{thebibliography}{99}
\bibitem{NoteResistDiff}Note that the resistivity $\rho$ is related to the corresponding
diffusion coefficient $D$ by relation $1/\rho=e^{2}\nu D$, where
$\nu$ is the normal DoS per spin.
\end{thebibliography}
\end{document}